\documentclass[conference]{IEEEtran}
\IEEEoverridecommandlockouts
\usepackage{amsmath,amsfonts}
\usepackage{algorithmic}
\usepackage{algorithm}
\usepackage{array}
\usepackage{textcomp}
\usepackage{stfloats}
\usepackage{subfigure}
\usepackage{url}
\usepackage{verbatim}
\usepackage{graphicx}
\usepackage{cite}
\usepackage{xcolor}
\usepackage{makecell} 

\def\BibTeX{{\rm B\kern-.05em{\sc i\kern-.025em b}\kern-.08em
    T\kern-.1667em\lower.7ex\hbox{E}\kern-.125emX}}
\begin{document}

\title{Stochastic Geometry Analysis of Asymmetric Uplink Interference for Urban UAV-RC Networks
\thanks{This work is supported in part by the NSF awards CNS-1939334 and CNS-2332835.}
}

\author{Donggu Lee$^1$, Sung Joon Maeng$^2$, and Ismail Guvenc$^1$ \\
$^1$Department of Electrical and Computer Engineering, North Carolina State University, Raleigh, NC, USA \\
$^2$Department of Electrical and Electronic Engineering, Hanyang University, Ansan, South Korea \\
E-mail: \{dlee42, iguvenc\}@ncsu.edu, sjmaeng@hanyang.ac.kr}

\markboth{Journal of \LaTeX\ Class Files,~Vol.~14, No.~8, August~2021}%
{Shell \MakeLowercase{\textit{et al.}}: A Sample Article Using IEEEtran.cls for IEEE Journals}


\maketitle

\begin{abstract}
Uncrewed aerial vehicles (UAVs) have emerged as a flexible platform for providing coverage over challenging environments, particularly for public safety and surveillance missions in urban areas. However, deploying the UAVs in dense urban areas introduces unique challenges, most notably asymmetric uplink (UL, remote controller to UAV) interference due to a higher chance of line-of-sight (LoS) interference at the UAV. In this letter, we propose a stochastic geometry framework to tractably analyze the large-scale asymmetric interference in urban areas. We incorporate a log-Gaussian Cox process (LGCP) model to capture the spatial correlation of the interference field in both UL and downlink (DL) as a function of the UAV altitude and the two-dimensional (2-D) distance between the remote controller and UAV. To quantify the UL and the DL interference asymmetry, we also define the interference asymmetry ratio characterizing the interference disparity between the UL and the DL. Our numerical results demonstrate that the interference asymmetry ratio increases as the UAV altitude and 2-D distance increase, highlighting that the UL interference worsens.
\end{abstract}


\section{Introduction}
\begin{figure}[t!]
	\centering
	\includegraphics[width=0.7\columnwidth]{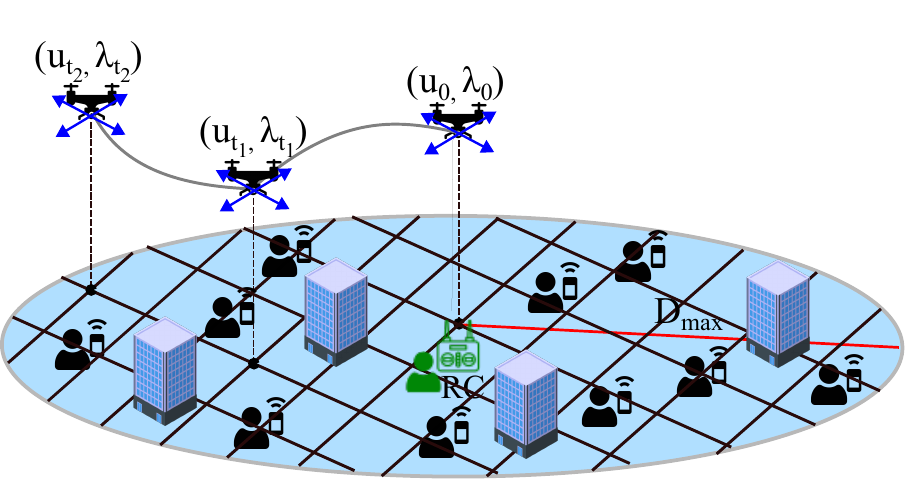}\vspace{-0.5mm}
	\caption{UAV RC network model with the RC located at the origin. The UAV flies over possible locations illustrated as a grid starting from the origin. Interferers are distributed within the coverage. The aggregate interference power in the UL (at the UAV) is generally stronger than the DL due to a higher chance of LoS interference from the ground.}\label{fig:ill_to_lgcp}
\end{figure}

\IEEEPARstart{U}{ncrewed} aerial vehicles (UAVs) provide flexible coverage over unfavorable and challenging environments~\cite{access_UAV_application}. Consequently, UAVs are increasingly adopted in metropolitan areas for public safety and surveillance purposes, e.g., drones as first responders (DFR)~\cite{skydio_dfr_best_practices_2024}. However, deploying UAVs in urban environments introduces unique challenges compared to conventional cellular communication systems~\cite{proc_ieee_UAV_application}, such as severe interference from terrestrial networks~\cite{asymmetry_UAV, asymmetry_uav_journal}, and line-of-sight (LoS) dominance of UAV-ground channels. In our previous works~\cite{OJCOMS_arxiv, MILCOM_arxiv}, interference in uplink (UL), i.e., remote controller (RC) on the ground to UAV in the air, was shown to play a critical role in UAV networks by severely degrading uplink ACK/NACK transmission. The UL is more vulnerable to the interference than the downlink (DL, UAV to RC) due to a higher chance of LoS interference in the UL, while the RC is comparatively isolated from the interference, as shown in Fig.~\ref{fig:ill_to_lgcp}. 

To investigate the interference over a large-scale area, stochastic geometry has been widely used as it facilitates a probabilistic analysis~\cite{stochastic_geometry_survey}. In~\cite{stochastic_related_work_1}, the coverage and spectral efficiency performance are investigated for UAV-based millimeter wave (mmWave) networks in urban areas. Interference has been analyzed using the Laplacian transform, where the distribution of the users on the ground is modeled as a typical Poisson cluster process (PCP). Leveraging this stochastic geometry model, an optimal altitude of the UAV is derived to maximize the coverage and spectral efficiency. 
\vspace{-0.5mm}
A similar study can be found in~\cite{stochastic_related_work_2}, which analyzes the coverage probability for UAV-aided post-disaster cellular networks. Here, terrestrial base stations and UAVs are distributed under an inhomogeneous Poisson point process (IPPP) and a uniform distribution model, respectively. Numerical analysis is carried out to evaluate the coverage probability performance with respect to various aspects of deployment scenarios, such as the distance between user equipments (UEs) and the disaster epicenter, the distance radius, and the quality of resilience.

In this letter, to tractably analyze the UL/DL interference asymmetry in urban scenarios, we propose, to the best of our knowledge, the first stochastic geometry framework addressing this problem, whereas our previous work~\cite{OJCOMS_arxiv, MILCOM_arxiv} relied on simulations and measurement campaigns. In addition to our previous work in~\cite{sungjoon_stochastic_paper}, where the UEs are distributed under a homogeneous Poisson point process (HPPP), we adopt a log-Gaussian Cox process (LGCP) formulation to capture the spatial correlation of interference fields at UL and DL with respect to the two-dimensional (2-D) distance between UAV and RC. Moreover, we also define the UL interference asymmetry ratio to quantify the asymmetry and to analyze the impact of the spatial correlation. Our numerical results show that the interference asymmetry ratio increases as the UAV altitude and 2-D distance increase, which demonstrates that UL interference worsens compared to the DL, by using the Monte-Carlo method and closed-form analysis derived from the LGCP-based stochastic geometry approach.



\section{System Model}\label{sec:system_model}

\subsection{Stochastic Geometry-based  Interferer Distribution with UAV Mobility}\label{sec:sto_geo}
In a homogeneous Poisson point process (HPPP)-based stochastic geometry framework, ground interferers are randomly distributed with a uniform node density. At the initial time of flight, the RC and UAV are co-located in the 2-D plane. Therefore, the interference fields observed at the UAV (UL, RC-to-UAV) and the RC (DL, UAV-to-RC) are statistically identical and highly correlated. However, as the UAV flies away from the RC, the interference fields observed at the UAV and RC become spatially uncorrelated. This indicates that the aggregate interference power in the UL and the DL is influenced by the UAV’s location.

To model the movement of the UAV, we adopt the Markov random walk model over a discretized 2-D grid, denoted by $\mathcal{S}=\{s_{0,0}, s_{0,1}, \cdots, s_{i,j}\}$. The RC is located at the center grid point (origin) $s_{0,0}$, where the UAV is also initially placed ($u_{0}=s_{0,0}$). The altitude of the UAV is fixed to $H$, and the node density of the interferers at the initial time is denoted as $\lambda_0$. The 2-D location of the UAV and the node density of interferers at the location of the UAV at time step $t$ are denoted $u_t=(x_t, y_t)$ and $\lambda_{t}$, respectively. At each time step, the UAV moves to one of its four neighboring grid points in the cardinal directions (north, east, south, west) with distance $\Delta$. We consider the uniform probability distribution for the movement direction of the UAV. Then, the transition probability from the current position to any of its adjacent grid points is given by
\begin{align}
    &\mathsf{Pr}\{u_{t+1}=(x_t \pm \Delta, y_t)|u_t\}  \nonumber \\ =~ &\mathsf{Pr}\{u_{t+1}=(x_t, y_t \pm \Delta)|u_t\} =\frac{1}{4}.
\end{align}
The maximum 2-D distance between the RC and the UAV is constrained by $D_{\max}$, considering the coverage of the connectivity.


\subsection{Log-Gaussian Cox Process for Spatial Correlation of the Node Density between UL and DL}
As discussed in Section~\ref{sec:sto_geo}, the interference fields observed at the UAV (UL) and the RC (DL) are correlated depending on the distance between them. To capture this spatial correlation, we adopt the LGCP for modeling the spatial variability of the interferer density in UL ($\lambda_{t,\mathrm{UL}}$) and DL ($\lambda_{t,\mathrm{DL}}$). In general, a Cox process is known as a “doubly stochastic Poisson process” where the node density is governed by an underlying stochastic process over space. LGCP is a special case of the Cox process, where the node density is modeled as the exponential of a Gaussian random process. In our formulation, the node density of the interferers at the UAV (UL) and RC (DL) is modeled by a Gaussian process with spatial correlation depending on the distance between the UAV and the RC.

The node density of interferers in UL ($\lambda_{t, \mathrm{ul}}$) and DL ($\lambda_{t, \mathrm{dl}}$) in our considered scenarios follows LGCP, expressed as
\begin{equation}\label{eq:lam_t}
    \boldsymbol{\lambda}_t\sim\exp{(\mathcal{GP}(\mu_0,\boldsymbol{C}(u_0,u_t)))},
\end{equation}
where $\boldsymbol{\lambda}_t = (\lambda_{t, \mathrm{ul}}, ~\lambda_{t, \mathrm{dl}})^T$, $\boldsymbol{C}(\cdot, \cdot)$, and $\mathcal{GP}(\cdot)$ indicate a node density vector, the covariance matrix, and a Gaussian process, respectively. We assume that the mean of the Gaussian process is constant as $\mu_0$, which implies that regardless of the location of the UAV, the expectation of the node densities of the interferers in both UL and DL are identical as follows: 
\begin{equation}\label{eq:mu_bar_relationship}
    \mathbb{E}[{\lambda_{t,{\mathrm{ul}}}}] = \mathbb{E}[\lambda_{t,{\mathrm{dl}}}] = \exp(\mu_0+\frac{1}{2}\sigma_0^2) = \bar{\lambda}_0,
\end{equation}
where $\sigma_0^2$ is the variance of the log-Gaussian process, i.e., $\mathrm{Var}[\ln(\lambda_{t, \mathrm{ul}})] = \mathrm{Var}[\ln(\lambda_{t, \mathrm{dl}})] = \sigma_0^2$. The covariance matrix $\boldsymbol{C}(\cdot, \cdot)$ can be written as
\begin{equation}\label{eq:covariance_matrix}
\mathbf{C}(u_0, u_t) = \sigma_0^2 
\begin{bmatrix}
1 & \xi(u_0, u_t) \\
\xi(u_0, u_t) & 1
\end{bmatrix}, 
\end{equation}
where the spatial correlation function $\xi(\cdot, \cdot)\in (0,1]$ is modeled as an exponential decay function with respect to the 2-D distance between the UAV and the RC $d(u_0, u_t)$ at time step $t$, which can be expressed as
\begin{equation}\label{eq:correlation}
    \xi(u_0, u_t) = \exp(-k_0d(u_0,u_t)),
\end{equation}
where $k_0$ is a configurable fitting parameter depending on the environment. In the LGCP formulation, the expectation of node density of the interferers $\bar{\lambda}_0$ is configured, and then the mean of the Gaussian process $\mu_0$ is obtained from \eqref{eq:mu_bar_relationship}, given by 
\begin{equation}
    \mu_0 = \ln(\bar{\lambda}_0) - \frac{1}{2}\sigma_0^2.
\end{equation}
The second-order moments of the node densities of the interferers in the UL and the DL can be derived as
\begin{align}
    \mathbb{E}\{(\lambda_{t,{\mathrm{ul}}})^2\}&=\mathbb{E}\{(\lambda_{t,{\mathrm{dl}}})^2\}=\bar{\lambda}_0^2\exp(\sigma^2_0),\label{eq:exp_lam_2m}\\
    \mathbb{E}\{\lambda_{t,{\mathrm{ul}}}\lambda_{t,{\mathrm{dl}}}\}&=\bar{\lambda}_0^2\exp{(\sigma^2_0\xi(u_0, u_t))}.\label{eq:exp_lam_2m_2}
\end{align}
From~\eqref{eq:mu_bar_relationship}, while the expectation of the node density of interferers in UL and DL is the same as $\bar{\lambda}_0$, interference asymmetry between UL and DL arises depending on the location of the UAV due to the spatial correlation of the node densities as captured by~\eqref{eq:covariance_matrix}.

\section{Aggregate Interference and Interference Ratio Asymmetry Analysis}\label{sec:aggregate_interference}

\subsection{LoS Probability}
In UAV-to-ground channel propagation, the probability of LoS connectivity increases as the UAV altitude increases. In \cite{LoS_prob_paper}, the altitude-dependent LoS probability is derived as
\begin{align}
    \mathrm{Pr}_{\rm los} = \Pi^{m}_{n=0}\left[1-\exp\left(-\frac{\left(H-\frac{(n+1/2){H}}{m+1}\right)^2}{2\gamma^2}\right)\right],\label{eq:2}
\end{align}
where $m=\text{floor}(r\sqrt{\delta\beta}-1)$. Although~\eqref{eq:2} can be closely approximated to a modified Sigmoid function (S-curve) with respect to the elevation angle $\theta$~\cite{LoS_prob_paper}, the approximated form is still not feasible to calculate the closed-form expression in~\eqref{eq:ul_exp1}. Alternatively, we approximate the LoS probability expression in~\eqref{eq:2} using a break-point exponential function formula as introduced in~\cite{sungjoon_stochastic_paper}
\begin{align}\label{eq:bp_2}
    \mathrm{Pr}_{\rm los} = \begin{cases}
    1 & {\rm R_0}<R<\kappa H\\
    \exp{(-\frac{\mu}H(R-\kappa H))} & R>\kappa H
    \end{cases}~,
\end{align}
where $\kappa H$ represents the break-point distance where the
probability function transitions from a constant value to
an exponential decay, and $\mu$, $\kappa$ are environmental fitting
parameters where $\mu = 0.625$,
$\kappa = 1.38$ were found to model urban environments, respectively in~\cite{sungjoon_stochastic_paper}. Note that, as shown in \cite[Fig.~2]{sungjoon_stochastic_paper}, the decaying slope of the approximated expression in \eqref{eq:bp_2} is highly close to the exact expression in \eqref{eq:2}.

\subsection{Expectation of Aggregate Interference Power in UL/DL}\label{sec:exp_ul}
The aggregate interference power of UL (RC-to-UAV) can be expressed as,
\begin{align}\label{eq:I_ul_def}              {I_{\mathrm{ul}}}=\sum_{r_i\in\Phi(\lambda_{t, \rm{ul}})}\frac{P_{t, \mathrm{I}}G_{\mathrm{I}}G_{\mathrm{uav}}c^2}{(4\pi)^2f^2R_i^{\eta_{\mathrm{x,ul}}}}h_i,
\end{align}
where $P_{t, \mathrm{I}}$, $G_{\mathrm{I}}$, $G_{\mathrm{uav}}$, $c$, and $f$ denote the transmit power of an interferer, the transmit antenna gain of an interferer, the receive antenna gain of the UAV, the speed of light, and the carrier frequency, respectively. Moreover, $h_i\sim \exp(1)$ represents Rayleigh small-scale fading, while $R_i$, $r_i$ are three-dimensional (3-D) distance and 2-D distance between UAV and $i$-th UE, where $R_i=\sqrt{r_i^2+{H}^2}$. The path loss exponent in the UL can be expressed by  $\eta_{\mathrm{x,ul}}$, $\mathrm{x}\in\{\mathrm{los, nlos}\}$ for LoS/NLoS conditions and the state of the A2G channel, respectively.

By using Campbell Theorem and \eqref{eq:mu_bar_relationship}, the expectation of aggregate received signal considering LoS probability $\mathrm{Pr}_{\mathrm{los}}(R)$ is given by
\begin{align}
    &\mathbb{E}\left\{I_{\mathrm{ul}}\right\}=\mathbb{E}\left\{\sum_{r_i\in\Phi(\lambda_{t, \rm{ul}})}\frac{P_{t, \mathrm{I}}G_{\mathrm{I}}G_{\mathrm{uav}}c^2}{(4\pi)^2f^2R_i^{\eta_{\mathrm{x, ul}}}}h_i\right\} \label{eq:ul_exp1} \\
   & \approx \beta_{\mathrm{UL}} \mathbb{E}  \left\{\lambda_{t,\rm{ul}}\right\}\mathbb{E}\left\{h_i\right\}\Big\{\int_{{R_0}}^{\infty} R^{-\eta_{\mathrm{los,ul}}+1}\mathrm{Pr}_{\rm los}(R)\mathrm{d}R \nonumber \\ &~~~~~~~~~~~~~~~~~~~~~~~+\int_{{R_0}}^{\infty} R^{-\eta_{\mathrm{nlos,ul}}+1}(1-\mathrm{Pr}_{\rm los}(R))\mathrm{d}R \Big\}\nonumber\\
   & = \beta_{\mathrm{UL}} \bar{\lambda}_0  \Big\{\int_{{R_0}}^{\infty} R^{-\eta_{\mathrm{los,ul}}+1}\mathrm{Pr}_{\rm los}(R)\mathrm{d}R \nonumber \\ & ~~~~~~~~~~~~~~~+\int_{{ R_0}}^{\infty} R^{-\eta_{\mathrm{nlos,ul}}+1}(1-\mathrm{Pr}_{\rm los}(R))\mathrm{d}R\Big\}, \label{eq:ul_exp2}
\end{align}
where $\beta_{\mathrm{UL}}=\{2\pi P_{t, \mathrm{I}}G_{\mathrm{I}}G_{\mathrm{uav}}c^2\}/\{(4\pi)^2f^2\}$, ${R_0}=\sqrt{ H^2+r_0^2}$, $\mathbb{E}\left\{h_i\right\}=1$, and $r_0$ is the minimum horizontal distance between the typical UAV and interferers. The approximation comes from transforming a uniform distribution over $r_i$ into $R_i$. By applying \eqref{eq:bp_2} to \eqref{eq:ul_exp2}, we can obtain the closed-form expression of the $\mathbb{E}\left\{I_{\mathrm{ul}}\right\}$ in~\eqref{eq:exp_ul_cl},
\begin{figure*}[t] 
\begin{align}\label{eq:exp_ul_cl}
    &\mathbb{E}\left\{I_{\mathrm{ul}}\right\}=\beta_{\rm UL}\bar{\lambda}_0
    \begin{cases}
    \log\left(\frac{\kappa{H}}{R_0}\right)+e^{\mu\kappa}\Gamma(0,\mu\kappa)-e^{\mu\kappa}\left(\frac{\mu}{H}\right)^{\eta_{\rm nlos, \rm ul}-2}\Gamma(2-\eta_{\rm nlos,ul},\mu\kappa)+\frac{(\kappa{H})^{-\eta_{\rm nlos,ul}+2}}{\eta_{\rm nlos,ul}-2},\quad\kappa{H}\geq{R_0}\\
         e^{\mu\kappa}\Gamma(0,\frac{\mu R_0}{H})-e^{\mu\kappa}\left(\frac{\mu}{H}\right)^{\eta_{\rm nlos, \rm ul}-2}\Gamma(2-\eta_{\rm nlos,ul},\frac{\mu R_0}{H})+\frac{(R_0)^{-\eta_{\rm nlos,ul}+2}}{\eta_{\rm nlos,ul}-2},\quad\kappa{H}<{R_0}
    \end{cases}
\end{align}
\hrulefill 
\end{figure*} 
where we assume that the pathloss exponent of LoS link is $\eta_{\mathrm{los,ul}}=2$, the pathloss exponent of NLoS link is $\eta_{\mathrm{nlos,ul}}>2$,  $\Gamma(s,a)=\int_{a}^{\infty}x^{s-1}\exp{(-x)}\mathrm{d}x$ denotes the upper incomplete gamma function, and we apply the property of $\int_{a}^{\infty}x^{s-1}\exp{(-bx)}\mathrm{d}x=\frac{1}{b^s}\Gamma(s,ba)$.

The aggregate interference power of DL (UAV-to-RC) can be expressed as
\begin{align}\label{eq:I_dl_def}
    I_{{\mathrm{ dl}}}=\sum_{r_i\in\Phi(\lambda_{t,\rm dl})}\frac{P_{t, \mathrm{I}}G_{\mathrm{I}}G_{\mathrm{rc}}c^2}{(4\pi)^2f^2r_i^{\eta_{\mathrm{nlos, dl}}}}h_i,
\end{align}
where $r_i$ and $G_{\mathrm{rc}}$ indicate the 2-D distance between the typical RC and the $i$-th interferer and the receive antenna gain of the RC. We assume that the interference signals in DL are in NLoS due to the blockage by the surrounding buildings. The closed-form expression for the expectation of the aggregate interference power in DL can be derived from \eqref{eq:mu_bar_relationship} as
\begin{align}
    \mathbb{E}\left\{I_{\mathrm{dl}}\right\}&=\mathbb{E}\left\{\sum_{r_i\in\Phi(\lambda_{t,\rm dl})}\frac{P_{t, \mathrm{I}}G_{\mathrm{I}}G_{\mathrm{rc}}c^2}{(4\pi)^2f^2r_i^{\eta_{\mathrm{nlos,dl}}}}h_i\right\} \label{eq:exp_dl_1} \\
    &= \beta_{\mathrm{DL}} \mathbb{E}\left\{\lambda_{t,\rm dl}\right\}\mathbb{E}\left\{h_i\right\}\int_{{ r_0}}^{\infty} r^{-\eta_{\mathrm{nlos,dl}}+1}\mathrm{d}r\nonumber\\
    &=\beta_{\mathrm{DL}} \bar{\lambda}_0 \left(\frac{r_0^{-\eta_{\mathrm{nlos,dl}}+2}}{\eta_{\mathrm{nlos,dl}}-2}\right), \label{eq:exp_dl_2}
\end{align}
where $\beta_{\mathrm{DL}}=\{2\pi P_{t, \mathrm{I}}G_{\mathrm{I}}G_{\mathrm{rc}}c^2\}/\{(4\pi)^2f^2\}$ and $r_0>1$~m is the minimum horizontal distance between the typical RC and interferers.

\subsection{Variance of Aggregate Interference Power in DL}\label{sec:var_dl}
The second moment of the aggregate interference power in DL can be derived from \eqref{eq:I_dl_def} as
\begin{align}\label{eq:I_dl_2m}
    &\mathbb{E}\{I_{\mathrm{dl}}^2\}=\mathbb{E}\left\{\sum_{r_i\in\Phi(\lambda_{t,\rm dl})}\left(\frac{\beta_{\mathrm{DL}}}{2\pi}\right)^2r_i^{-2\eta_{\mathrm{nlos,dl}}}\right\}\mathbb{E}\{h_i^2\}+\nonumber\\
    &\mathbb{E}\left\{\sum_{r_i,r_j\in\Phi(\lambda_{t,\rm dl})}^{i\neq j}\left(\frac{\beta_{\mathrm{DL}}}{2\pi}\right)^2r_i^{-\eta_{\mathrm{nlos,dl}}}r_j^{-\eta_{\mathrm{nlos,dl}}}\right\}\mathbb{E}\{h_i\}\mathbb{E}\{h_j\}\nonumber\\ 
    &=\frac{\beta_{\rm DL}^2}{\pi}\mathbb{E}\{\lambda_{t,\rm dl}\}\int_{{ r_0}}^{\infty} r^{1-2\eta_{\mathrm{nlos,dl}}}\mathrm{d}r\nonumber\\
    &+\beta_{\rm DL}^2\mathbb{E}\{(\lambda_{t,\rm dl})^2\}\int_{{ r_0}}^{\infty} r^{1-\eta_{\mathrm{nlos,dl}}}\mathrm{d}r\int_{{ r_0}}^{\infty} r^{1-\eta_{\mathrm{nlos,dl}}}\mathrm{d}r,
\end{align}
where $\mathbb{E}\{h_i^2\}=2$. Then, by using \eqref{eq:exp_lam_2m}, \eqref{eq:exp_dl_2}, \eqref{eq:I_dl_2m}, the closed-form expression of the variance of the aggregate interference power in DL can be written as
\begin{align}\label{eq:var_dl}
    &\mathrm{Var}\{I_{\mathrm{dl}}\} = \mathbb{E}\{I_{\mathrm{dl}}^2\} - \mathbb{E}\{I_{\mathrm{dl}}\}^2 = \frac{\beta_{\mathrm{DL}}^2\bar{\lambda}_0}{\pi}\Big(\frac{r_0^{2-2\eta_{\mathrm{nlos,dl}}}}{2\eta_{\mathrm{nlos,dl}}-2}\Big)  \nonumber \\
    &~~~~~~~~~~~~ +\beta_{\mathrm{DL}}^2 \big\{\bar{\lambda}_0^2\big(\exp(\sigma_0^2)-1 \big) \big\} \Big(\frac{r_0^{-\eta_{\mathrm{nlos,dl}}+2}}{\eta_{\mathrm{nlos,dl}}-2} \Big)^2.
\end{align}

\subsection{Mean Product of Aggregate Interference Power}\label{sec:exp_ul_dl}
From \eqref{eq:exp_lam_2m_2}, \eqref{eq:I_ul_def}, \eqref{eq:I_dl_def}, the closed-form expression of the mean product of aggregate interference power of UL and DL can be expressed as 
\begin{align}\label{eq:exp_ul_dl}
    &\mathbb{E}\{I_{\mathrm{ul}}I_{\mathrm{dl}} \}= \beta_{\mathrm{UL}}\beta_{\mathrm{DL}}\mathbb{E}\{\lambda_{t, \mathrm{ul}} \lambda_{t, \mathrm{dl}} \} \mathbb{E}\{h_i\}\mathbb{E}\{h_j\} \nonumber \\
    &~~~~~~~~~~~~~~~~~~~\Bigg\{ \int_{R_0}^{\infty}\mathbb{E}\{R^{-\eta_{\mathrm{x, ul}}}\} \mathrm{d}R \times \int_{r_0}^{\infty}r^{-\eta_{\mathrm{nlos,dl}}} \mathrm{d}r \Bigg\} \nonumber \\
    &=\beta_{\mathrm{UL}}\beta_{\mathrm{DL}} \bar{\lambda}_0^2 \exp(\sigma_0^2 \xi(u_0,u_t))\Bigg[ \int_{R_0}^{\infty}\int_{r_0}^{\infty} r^{-\eta_{\mathrm{nlos,dl}}} \times \nonumber \\
    & \Big\{R^{-\eta_{\mathrm{los, ul}}+1}\mathrm{Pr_{los}}(R)+R^{-\eta_{\mathrm{nlos,ul}}+1}(1-\mathrm{Pr_{los}}(R)) \Big\}\mathrm{d}r\mathrm{d}R \Bigg]\nonumber\\
    &=\beta_{\mathrm{UL}}\beta_{\mathrm{DL}} \bar{\lambda}_0^2 \exp(\sigma_0^2 \xi(u_0,u_t))\left\{\frac{r_0^{-\eta_{\mathrm{nlos,dl}}+2}}{\eta_{\mathrm{nlos,dl}}-2}\right\}\times\nonumber\\
    &\left\{\log\left(\frac{\kappa{\rm H}}{R_0}\right)+e^{{\mu\kappa}}\Gamma(0,\mu\kappa)-\right.\nonumber\\
    &\left.e^{\mu\kappa}\left(\frac{\mu}{\rm H}\right)^{\eta_{\mathrm{nlos,ul}}-2}\Gamma(2-\eta_{\mathrm{nlos,ul}},\mu\kappa)+\frac{(\kappa{\rm H})^{-\eta_{\mathrm{nlos,ul}}+2}}{\eta_{\mathrm{nlos,ul}}-2}\right\}.
\end{align}

\subsection{UL Interference Asymmetry Ratio}
Due to the impact of the spatial correlation of the node density between UL and DL, the interference ratio of UL and DL varies with the location of the UAV. To analyze the variation in UL and DL interference asymmetry, we derive the closed-form expression of the expectation of the interference power ratio by applying the first and the second order moment statistics obtained in Section~\ref{sec:exp_ul} -- \ref{sec:exp_ul_dl}. 
The expectation of the UL interference asymmetry ratio can be defined as
\begin{equation}\label{eq:rho_monte_carlo}
    \rho_I=\mathbb{E}\left\{\frac{I_{\mathrm{ul}}}{I_{\mathrm{dl}}} \right\}.
\end{equation}
By using Taylor series expansion for the moments of ratio functions~\cite{van2000mean}, \eqref{eq:rho_monte_carlo} can be approximated as
\begin{align}\label{eq:rho_closed_form}
    \rho_{I}&\approx\frac{\mathbb{E}\left\{I_{\mathrm{ul}}\right\}}{\mathbb{E}\left\{I_{\mathrm{dl}}\right\}}+\frac{\mathbb{E}\left\{I_{\mathrm{ul}}\right\}\mathrm{Var}\left\{I_{\mathrm{dl}}\right\}}{\mathbb{E}\left\{I_{\mathrm{dl}}\right\}^3}-\frac{\mathrm{Cov}(I_{\mathrm{ul}},I_{\mathrm{dl}})}{\mathbb{E}\left\{I_{\mathrm{dl}}\right\}^2},
\end{align}
where the covariance between the aggregate interference power in the UL and the DL is denoted as $ \mathrm{Cov}(I_{\mathrm{ul}}, I_{\mathrm{dl}})=\mathbb{E}\{I_{\mathrm{ul}}I_{\mathrm{dl}} \}-\mathbb{E}\{I_{\mathrm{ul}} \}\mathbb{E}\{I_{\mathrm{dl}} \}$. By substituting  \eqref{eq:exp_ul_cl}, \eqref{eq:exp_dl_2}, \eqref{eq:I_dl_2m}, \eqref{eq:var_dl}, \eqref{eq:exp_ul_dl} into \eqref{eq:rho_closed_form}, the closed-form expression for the expectation of the UL interference asymmetry ratio can be obtained.

From the closed-form expression in \eqref{eq:rho_closed_form}, we have the following observations: 1) only the last term of the expression contains the covariance $\mathrm{Cov}(I_{\mathrm{ul}}, I_{\mathrm{dl}})$, which is governed by the spatial correlation $\xi(u_0, u_t)$ as a function of the distance between the RC and the UAV $d(u_0,u_t)$; 2) when the spatial correlation $\xi(u_0, u_t)$ becomes zero, this term becomes one ($\mathrm{Cov}(I_{\mathrm{ul}}, I_{\mathrm{dl}})=1$), which means that \eqref{eq:rho_closed_form} is maximized with respect to the distance $d(u_0,u_t)$; 3) as the distance $d(u_0,u_t)$ decreases, the spatial correlation $\xi(u_0, u_t)$ increases, and the third term with negative sign gradually increases, thereby reducing \eqref{eq:rho_closed_form} until the distance and the spatial correlation becomes $d(u_0,u_t)=0, \xi(u_0, u_t)=1$. In summary, \eqref{eq:rho_closed_form} gradually increases with $d(u_0,u_t)$, and converges to a constant value as $\xi(u_0, u_t)\to0$.

\section{Numerical Analysis}\label{sec:numerical_analysis}

\begin{table}[t!]
\centering
\caption{Simulation Parameters.}\vspace{-0.4mm}

\begin{tabular}{|c|c|c|}
\hline
\textbf{Parameter} & \textbf{Description} & \textbf{Value [unit]} \\ \hline
$r_{0}$ & \makecell{Minimum horizontal distance \\ between UAV and UEs} & $20$ [m] \\ \hline
$P_{t, \mathrm{I}}$ & Transmit power of interferers & $0.1$ [W] \\ \hline
$P_{t, \mathrm{rc}}$ & Transmit power of RC& $0.1$ [W] \\ \hline
$P_{t, \mathrm{uav}}$ & Transmit power of UAV & $0.1$ [W] \\ \hline
$G_{\mathrm{I}}$ & Antenna gain of interferers & $3$ [dBi] \\ \hline
$G_{\mathrm{rc}}$ & Antenna gain of RC & $6$ [dBi] \\ \hline
$G_{\mathrm{uav}}$ & Antenna gain of UAV & $6$ [dBi] \\ \hline
$\eta_{\mathrm{los,ul}}$ & Path loss coefficient for LoS UL & $2$ \\ \hline
$\eta_{\mathrm{nlos,ul}}$ & Path loss coefficient for NLoS UL & $3$ \\ \hline
$\eta_{\mathrm{nlos,dl}}$ & Path loss coefficient for NLoS DL & $3$ \\ \hline
$f$ & Carrier frequency & $3.5$ [GHz] \\ \hline
$\bar{\lambda}_0$ & Node densities of the interferers & $0.02$ [nodes/${\rm m}^2$] \\ \hline
$r_{\max}$ & Maximum radius of the environments & $3,000$ [m] \\ \hline
\end{tabular}\label{tab:parameters}\vspace{-0.2mm}
\end{table}

The simulation parameters for the urban UAV scenarios are summarized in Table~\ref{tab:parameters}. In this study, we set two different cases: 1) fixed 2-D distance with variable UAV altitude, and 2) fixed UAV altitude with variable 2-D distance. In both scenarios, we adopt a Monte-Carlo method-based approach for simulations and closed-form analysis.



\begin{figure}[t!]
    \centering 
    \subfigure[$\rho_I$ versus altitude with the fixed 2-D distance of $30$~m.]{    \includegraphics[trim={0.4cm, 0, 1.0cm, 0.35cm},clip, width=0.8\columnwidth]{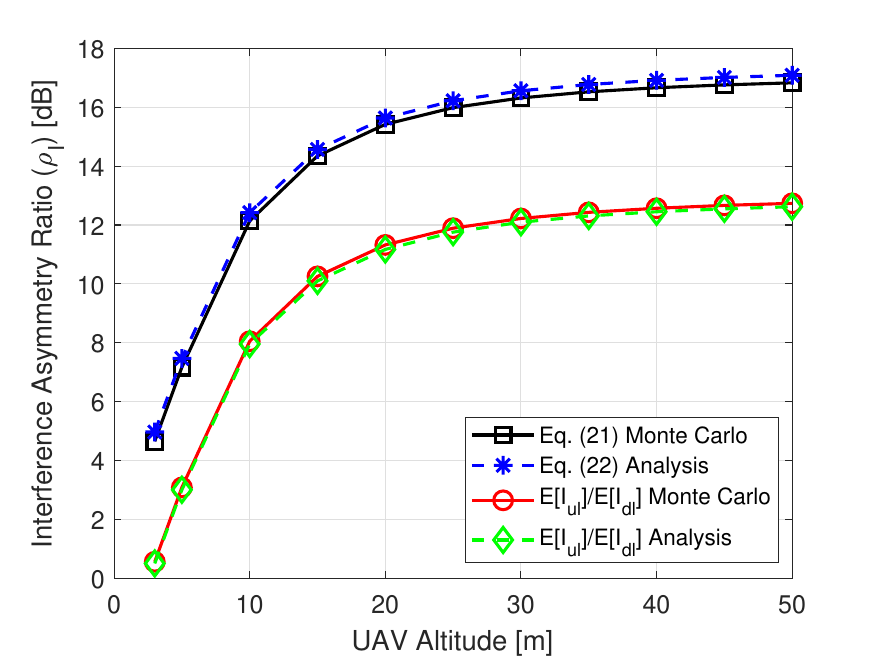}
    \label{fig:rho_altitude}}\vspace{-0.3mm}
    \subfigure[$\rho_I$ versus 2-D distance with the fixed UAV altitude of $30$~m.]{    \includegraphics[trim={0.1cm, 0, 1.0cm, 0.35cm},clip, width=0.8\columnwidth]{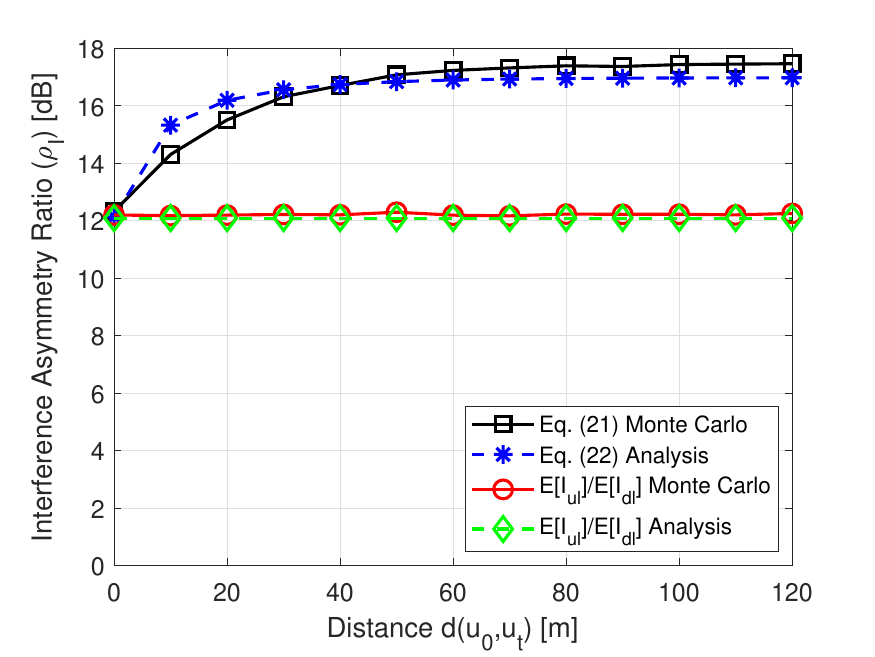}
    \label{fig:rho_2D_distance}}
    \caption{$\rho_I$ over altitude and 2-D distance results in urban scenarios.}
    \label{fig:rho_altitude_2Ddistance}\vspace{-0.3mm}
\end{figure}

\begin{figure}[t!]
    \centering 
    \subfigure[$\mathrm{Cov}(I_{\mathrm{ul}},I_{\mathrm{dl}})$ versus altitude.]{    \includegraphics[trim={0.4cm, 0, 1.0cm, 0.2cm},clip, width=0.8\columnwidth]{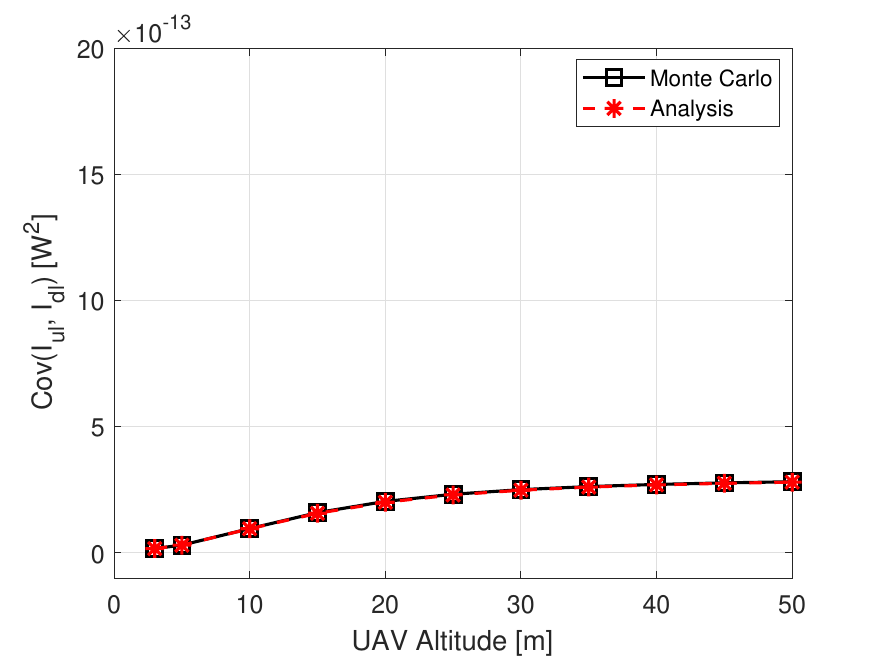}
    \label{fig:cov_altitude}}\vspace{-0.3mm}
    \subfigure[$\mathrm{Cov}(I_{\mathrm{ul}},I_{\mathrm{dl}})$ versus 2-D distance.]{    \includegraphics[trim={0.1cm, 0, 1.0cm, 0.2cm},clip, width=0.8\columnwidth]{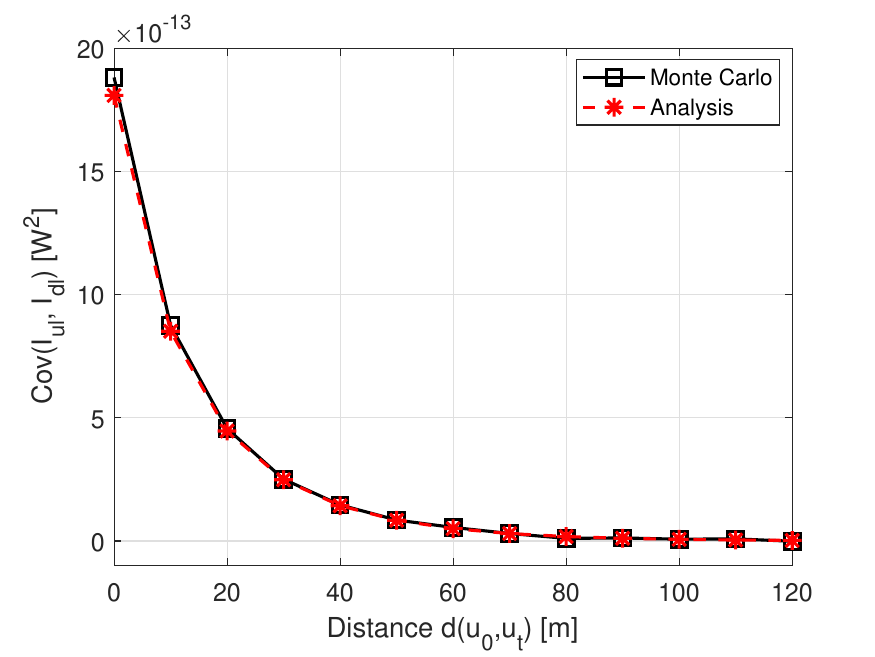}
    \label{fig:cov_2D_distance}}\vspace{-0.3mm}
    \caption{$\mathrm{Cov}(I_{\mathrm{ul}},I_{\mathrm{dl}})$ over altitude and 2-D distance results in urban scenarios.}
    \label{fig:cov_altitude_2Ddistance}\vspace{-0.5mm}
\end{figure}

Fig.~\ref{fig:rho_altitude_2Ddistance} shows the simulation results for $\rho_I$ based on~\eqref{eq:rho_monte_carlo} and~\eqref{eq:rho_closed_form} for Monte-Carlo and closed-form approaches with the fixed 2-D distance and altitude, respectively. In addition to $\rho_I$, the results for $\mathbb{E}\{I_{\mathrm{ul}}\}_{\mathrm{dB}}-\mathbb{E}\{I_{\mathrm{dl}}\}_{\mathrm{dB}}=\mathbb{E}\{I_{\mathrm{ul}}\}/\mathbb{E}\{I_{\mathrm{dl}}\}$, for both Monte-Carlo and closed-form analysis are also included for comparison purposes. As seen in Fig.~\ref{fig:rho_altitude}, both closed-form and Monte-Carlo analyses tend to increase as the altitude of the UAV increases. This indicates that the interference at the UL becomes more dominant than the DL as the UAV altitude increases. A gap of approximately $4$~dB between $\rho_I$ and $\mathbb{E}\{I_{\mathrm{ul}}\}/\mathbb{E}\{I_{\mathrm{dl}}\}$ is observed, which originates from the second and the last terms in~\eqref{eq:rho_closed_form}. Furthermore, the covariance of interference in the UL and the DL $\mathrm{Cov}(I_{\mathrm{ul}}, I_{\mathrm{dl}})$ results over UAV altitude are shown in Fig.~\ref{fig:cov_altitude}. The covariance of UL and DL interference shows a slight increase with the UAV altitude. However, since the range of variation of $\mathbb{E}\{I_{\mathrm{ul}}\}$ is larger than $\mathrm{Cov}(I_{\mathrm{ul}}, I_{\mathrm{dl}})$, $\rho_I$ is increased as the UAV altitude increases. 

In Fig.~\ref{fig:rho_2D_distance}, numerical results of $\rho_I$ over variable 2-D distance $d(u_0, u_t)$ are presented, where $\rho_I$ increases within a relatively narrow range compared to the variable UAV altitude case, as the 2-D distance increases. In contrast to the altitude-dependent results, $\mathbb{E}\{I_{\mathrm{ul}}\}/\mathbb{E}\{I_{\mathrm{dl}}\}$ remains steady because the node densities of the UL and the DL share the same mean and variance regardless of the UAV location, resulting in steady variations of both $\mathbb{E}\{I_{\mathrm{ul}}\}$ and $\mathbb{E}\{I_{\mathrm{dl}}\}$. In Fig.~\ref{fig:cov_2D_distance}, the covariance in the UL and the DL interference over 2-D distance is shown. Compared to the variable UAV altitude cases, it shows an exponential decreasing pattern as a function of 2-D distance. This stems from the correlation term $\xi(u_0,u_t)$ defined in~\eqref{eq:correlation}, which is modeled as an exponentially decreasing function of the 2-D distance between the UAV and the RC. Since this $\xi(u_0,u_t)$ appears in the joint expectation term $\mathbb{E}\{I_{\mathrm{ul}}I_{\mathrm{dl}} \}$ in~\eqref{eq:exp_ul_dl}, the covariance, i.e., $ \mathrm{Cov}(I_{\mathrm{ul}}, I_{\mathrm{dl}})=\mathbb{E}\{I_{\mathrm{ul}}I_{\mathrm{dl}} \}-\mathbb{E}\{I_{\mathrm{ul}} \}\mathbb{E}\{I_{\mathrm{dl}} \}$, decreases as the 2-D distance increases. Moreover, the decrease in $\mathrm{Cov}(I_{\mathrm{ul}}, I_{\mathrm{dl}})$ contributes to the increase in $\rho_I$, as in Fig.~\ref{fig:rho_2D_distance} by reducing the magnitude of the negative term in~\eqref{eq:rho_closed_form}, while other terms remain steady.

\section{Conclusion}\label{sec:conclusion}
In this letter, we adopted an LGCP-based stochastic geometry framework to tractably analyze the UL interference asymmetry in large-scale urban UAV networks. We also defined the interference asymmetry ratio to quantify the UL interference asymmetry. We used the Monte-Carlo method and closed-form approaches in our numerical analysis. As shown in our numerical results, the interference asymmetry ratio increased as the UAV altitude and 2-D distance between RC and UAV increased. Thus, it is essential to maintain a balanced link condition in the UL and the DL to fully employ its capacity in the UAV networks.

\bibliographystyle{IEEEtran}
\bibliography{ref}

\end{document}